\documentclass[twocolumn,amsmath,amssymb,pra,superscriptaddress]{revtex4-1}
\usepackage[dvipdfmx]{graphicx}
\usepackage{dcolumn}
\usepackage{bm}
\usepackage{subfigure}
\usepackage{booktabs}
\usepackage{color}
\newcounter{one}
\def\ket#1{\mbox{\boldmath $#1$}}
\newcommand{\bracket}[1]{\left\langle #1 \right\rangle}

\newcommand{\affA}{
Artificial Intelligence Research Center, 
National Institute of Advanced Industrial Science and Technology, 
2-3-26 Aomi, Koto-ku, Tokyo, Japan
}

\begin{document}

\title{\textbf{Algorithmic infeasibility of community detection in higher-order networks}}

\author{Tatsuro Kawamoto}
\affiliation{\affA}

\begin{abstract}
In principle, higher-order networks that have multiple edge types are more informative than their lower-order counterparts. 
In practice, however, excessively rich information may be algorithmically infeasible to extract. 
It requires an algorithm that assumes a high-dimensional model and such an algorithm may perform poorly or be extremely sensitive to the initial estimate of the model parameters. 
Herein, we address this problem of community detection through a detectability analysis. 
We focus on the expectation--maximization (EM) algorithm with belief propagation (BP), and analytically derive its algorithmic detectability threshold, i.e., the limit of the modular structure strength below which the algorithm can no longer detect any modular structures. 
The results indicate the existence of a phase in which the community detection of a lower-order network outperforms its higher-order counterpart. 
\end{abstract}
 
\maketitle

Suppose we have a network, or a graph, in which the edges represent the similarities among the vertices. 
A community detection algorithm partitions the graph into subgraphs such that the vertices in the same group are similar to each other. 
A classical example of this is the detection of social groups in a social network, in which the vertices and edges represent persons and friendships, respectively. 
Although the amount of research aiming to optimize an objective function based on the edge density is voluminous \cite{Fortunato2010,Fortunato2016,Leger2013}, approaches based on statistical inference, which is often referred to as blockmodeling \cite{Goldenberg2010,Peixoto2017tutorial}, have also been actively developed. 
The latter expects that each vertex in a network belongs to a module (or modules), and that the vertices in the same module have statistically equivalent connection patterns, and thus we execute an algorithm to infer the most likely module assignments. 

Suppose we collect some friendship data among a set of people based on interviews to build a social network. 
We can simply collect binary information (friend or not) regarding friendship, such as Zachary's \textit{karate club network} \cite{karateclub}. 
If we can collect a sufficient number of edges and the network exhibits a clear community structure, then detecting the social groups will not be too difficult. 
However, when it is difficult to collect a sufficient number of edges or when the community structure is too weak, we may be unable to detect a statistically significant structure; in this case, the network is considered to be in an undetectable phase. 
One possible way to overcome this issue is to collect finer information. 
For example, we can classify a relationship as friendly, acrimonious, or unknown; here, it can be regarded that the non-edges are resolved into the edges of a new type (enemy), and into non-edges (unknown), and thus the total number of edges increases. 
A classical example of a social network with richer information is Sampson's \textit{monk network} \cite{WangWong87}. 

If we add edges of yet another type, which may be qualitatively different from a friendship, such as esteem, the network, in principle, becomes more informative. 
Assuming that all types of edges indicate the same modular structure, the question becomes whether it is always better to utilize all of these edges. 
In other words, is it always better to analyze higher-order networks than their lower-order counterparts? 
In practice, the answer is no. 
A higher-order model that takes into account excessively rich information may be algorithmically infeasible, i.e., the algorithm will exhibit a poor performance. 

Although the algorithmic infeasibility is a fundamental practical problem in community detection, to the best of our knowledge, it has not been theoretically analyzed. 
Herein, we address this problem by focusing on the expectation--maximization (EM) algorithm. 
We establish the emergence of the infeasibility through the detectability analysis described below.

We consider networks with labeled edges, i.e., an element of the adjacency matrix $\ket{A}$ takes one of the $p+1$ values, $A_{ij} = \alpha \in \{0, 1, \dots, p\}$. 
We denote the set of vertices as $V$ ($|V| = N$) and the set of $\alpha$-edges as $E_{\alpha}$ ($\cup_{\alpha=1}^{p} E_{\alpha} = E$, $\lvert E_{\alpha}\rvert = L_{\alpha}$, and $\sum_{\alpha=1}^{p}L_{\alpha} = L$); edges with $\alpha=0$ represent non-edges. 
We denote an undirected edge between vertices $i$ and $j$ as $(i,j)$. 
Although we can generalize the analysis here to general modular structures with an arbitrary number of modules \cite{ADT-Full}, we focus on the community (i.e., assortative) structure of two equally sized modules for brevity. 
An analysis of directed networks is out of the scope of the current study. 

For statistical inference of a community structure, it is common to use a family of random graph models with a modular structure. 
In particular, the so-called stochastic blockmodel (SBM) \cite{holland1983stochastic,WangWong87,KarrerNewman2011} is a canonical model, and the SBM with labeled edges is called the labeled stochastic blockmodel (labeled SBM). 
When the sizes of the modules are equal, the labeled SBM is generated as follows. 
First, the module assignment $\sigma_{i} \in \{1,2\}$ $(i \in V)$ for each vertex is determined uniformly and randomly. 
Then, for each element of the adjacency matrix $\ket{A}$, based on the module assignments, $A_{ij} = \alpha \in \{0, \dots, p\}$ is assigned independently with probability $c^{\alpha}_{\sigma_{i}\sigma_{j}}/N$. 
Therefore, the likelihood function of the labeled SBM is 
\begin{align}
p(A, \ket{\sigma} \lvert \ket{c}^{\alpha}) 
&\propto \prod_{i<j} \prod_{\alpha = 0}^{p} \left( \frac{c^{\alpha}_{\sigma_{i}\sigma_{j}}}{N} \right)^{\delta_{A_{ij},\alpha}}. \label{labeledSBMlikelihood}
\end{align}
The matrix $\ket{c}^{\alpha}$ is a $2 \times 2$ matrix that determines the connection density of $\alpha$-edges within/between modules. 
This is called an affinity matrix and is subject to a normalization constraint $N^{-1} \sum_{\alpha=0}^{p} c^{\alpha}_{\sigma \sigma^{\prime}} = 1$ for any $\sigma$ and $\sigma^{\prime}$. 
For each $\alpha$, the affinity matrix element is parameterized by $c^{\alpha}_{11} = c^{\alpha}_{22} = c^{\alpha}_{\mathrm{in}}$ and $c^{\alpha}_{12} = c^{\alpha}_{21} = c^{\alpha}_{\mathrm{out}}$. 
We denote the average degree of $\alpha$-edges as $c_{\alpha}$ and $c = \sum_{\alpha>0} c_{\alpha}$. 
We also denote the fraction of $\alpha$-edges as $P_{\alpha}$, i.e., $P_{\alpha} = c_{\alpha}/c$.
Herein, we focus on the sparse regime of the SBM, i.e., $c_{\alpha} = O(1)$ for all $\alpha>0$. 

In the labeled SBM, the affinity matrices $\{ \ket{c}^{\alpha} \}$ are the model parameters to be learned, and the module assignments $\{\sigma_{i}\}$ are the latent variables to be inferred. 
The model parameters are learned such that the marginalized log-likelihood $\log p(A \lvert \ket{c}^{\alpha}) = \log\sum_{\ket{\sigma}}p(A, \ket{\sigma} \lvert \ket{c}^{\alpha})$ of Eq.~(\ref{labeledSBMlikelihood}) is maximized, and the module-assignment distribution is obtained as the posterior estimate $p(\ket{\sigma} \lvert A, \ket{c}^{\alpha})$. 
Owing to the computational demand, however, an approximate algorithm is required. 

We can describe the EM algorithm as follows. 
Using the variational expression, $\log p(A \lvert \ket{c}^{\alpha})$ can be expressed as 
\begin{align}
\log p(A \lvert \ket{c}^{\alpha}) 
&= \mathbb{E}_{\psi}\left[\log \frac{p(A, \ket{\sigma} \lvert \ket{c}^{\alpha})}{\psi(\ket{\sigma})}\right] \notag\\
&\hspace{40pt}+ D_{\mathrm{KL}}\left( \psi(\ket{\sigma}) || p(\ket{\sigma} \lvert A, \ket{c}^{\alpha})\right), \label{VariationalExpression}
\end{align}
where $\psi(\ket{\sigma})$ is the variational probability distribution of the set of module assignments; $\mathbb{E}_{\psi}[\cdots]$ is the corresponding average; and $D_{\mathrm{KL}}(P || Q)$ represents the Kullback--Leibler divergence of distributions $P$ and $Q$. 
Equation (\ref{VariationalExpression}) indicates that, if $\psi(\ket{\sigma}) = p(\ket{\sigma} \lvert A, \ket{c}^{\alpha})$, $\log p(A \lvert \ket{c}^{\alpha})$ can be maximized by maximizing $\mathbb{E}_{\psi}\left[\log p(A, \ket{\sigma} \lvert \ket{c}^{\alpha})\right]$. 
Note that this is a double-optimization problem because $p(\ket{\sigma} \lvert A, \ket{c}^{\alpha})$ is conditioned on the estimate of $\ket{c}^{\alpha}$. 
For this reason, the EM algorithm iteratively updates the module-assignment inference (E-step) and the model-parameter learning (M-step). 

Because the exact computation of $p(\ket{\sigma} \lvert A, \ket{c}^{\alpha})$ is also computationally hard, we need another approximation during the E-step. 
Herein, we use and analyze the behavior of belief propagation (BP) \cite{Yedidia2001,Yedidia2003,MezardMontanari2009}. 
With BP, we compute the incomplete marginal probabilities, which we denote as $\{ \psi^{i \to j}_{\sigma_{i}} \}$ ($i,j \in V$); $\psi^{i \to j}_{\sigma_{i}}$ is the marginal distribution of vertex $i$ with missing knowledge of the edge between vertices $i$ and $j$. 
In the case of the labeled SBM, $\psi^{i \to j}_{\sigma_{i}}$ is iteratively updated as 
\begin{align}
\psi^{i \to j}_{\sigma_{i}} 
&\propto \prod_{\alpha>0} \mathrm{e}^{-\sum_{\ell} \sum_{\sigma_{\ell}} \psi^{\ell}_{\sigma_{\ell}} \hat{c}^{\alpha}_{\sigma_{\ell}\sigma}} 
\prod_{k \in \left[\partial i \backslash j\right]^{\alpha}} \left( \sum_{\sigma_{k}} \psi^{k \to i}_{\sigma_{k}} \hat{c}^{\alpha}_{\sigma_{k}\sigma} \right), \label{BP-1}
\end{align}
where $k \in \left[\partial i \backslash j\right]^{\alpha}$ is a neighbor of $i$ such that $(i,k) \in E_{\alpha}$ and $k \ne j$. 
By replacing $\left[\partial i \backslash j\right]^{\alpha}$ in Eq.~(\ref{BP-1}) with the neighboring vertices $\left[\partial i\right]^{\alpha}$ connected through $\alpha$-edges, we obtain the estimate of the complete marginal $\psi^{i}_{\sigma_{i}}$, 
\begin{align}
\psi^{i}_{\sigma_{i}} 
&\propto \prod_{\alpha>0} \mathrm{e}^{-\sum_{\ell} \sum_{\sigma_{\ell}} \psi^{\ell}_{\sigma_{\ell}} \hat{c}^{\alpha}_{\sigma_{\ell}\sigma}} 
\prod_{k \in \left[\partial i\right]^{\alpha}} \left( \sum_{\sigma_{k}} \psi^{k \to i}_{\sigma_{k}} \hat{c}^{\alpha}_{\sigma_{k}\sigma} \right). \label{BP-2}
\end{align}
The equations above are straightforward extensions of the BP equations for the standard SBM \cite{Decelle2011a}. 
BP is based on the tree approximation and is expected to be very accurate because we consider sparse networks. 

During the M-step, the model-parameter estimates $\{ \hat{c}^{\alpha}_{\mathrm{in}}, \hat{c}^{\alpha}_{\mathrm{out}} \}$ are updated as $\mathrm{argmax}\,\mathbb{E}_{\psi}\left[\log p(A, \ket{\sigma} \lvert \ket{c}^{\alpha})\right]$. 
Note here that $\hat{c}^{\alpha}_{\mathrm{in}}$ and $\hat{c}^{\alpha}_{\mathrm{out}}$ are not independent of each other because they are related to the $\alpha$-edge average degree $c_{\alpha}$ through $c_{\alpha} = (\hat{c}^{\alpha}_{\mathrm{in}} + \hat{c}^{\alpha}_{\mathrm{out}})/2$. 
Thus, the SBM is often characterized by $\Delta c_{\alpha} = c^{\alpha}_{\mathrm{in}} - c^{\alpha}_{\mathrm{out}}$, which can be interpreted as the strength of the community structure. 
Furthermore, owing to the constraints $c^{\alpha}_{\mathrm{in}} \ge 0$ and $c^{\alpha}_{\mathrm{out}} \ge 0$, we have $-2c_{\alpha} \le \Delta c_{\alpha} \le 2c_{\alpha}$. 
Hence, we also use the normalized parametrization $x_{\alpha} = 1/2 + \Delta c_{\alpha}/(4 c_{\alpha})$ such that $0 \le x_{\alpha} \le 1$. 
Using the latest estimates of $\{ \psi^{i \to j}_{\sigma_{i}} \}$, the normalized estimate $\hat{x}_{\alpha}$ of the strength of the community structure is updated as 
\begin{align}
\hat{x}^{(t+1)}_{\alpha} 
&= \hat{x}^{(t)}_{\alpha} \bracket{ \frac{1+2\left( X^{ij} - 1/2 \right)}{1 + 4(\hat{x}^{(t)}_{\alpha}-1/2)\left(X^{ij} - 1/2\right) } }_{E_{\alpha}}, \label{xUpdate}
\end{align}
where the superscript $(t)$ represents the $t$th update, $X^{ij} \equiv \sum_{\sigma}\psi^{i \to j}_{\sigma}\psi^{j \to i}_{\sigma}$, and $\bracket{Y_{ij}}_{E_{\alpha}} \equiv L_{\alpha}^{-1} \sum_{(i,j) \in E_{\alpha}} Y_{ij}$. 


\begin{figure}[t]
 \begin{center}
   \includegraphics[width=0.9 \columnwidth]{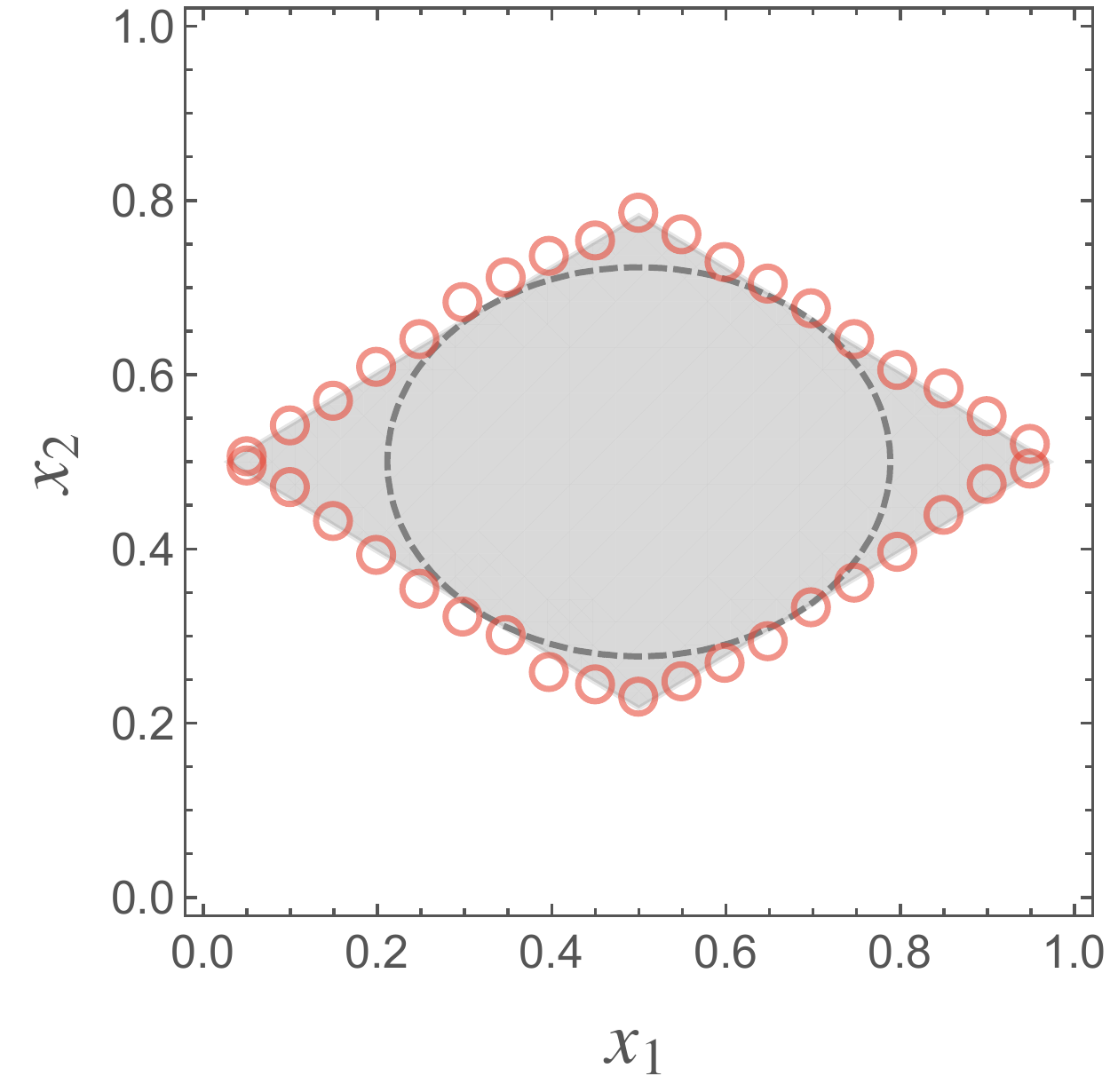}
 \end{center}
 \caption{
	(Color online) Detectability phase-diagram of the labeled SBM. 
	Each axis represents the normalized strength of the modular structure $x_{\alpha}$ ($\alpha \in \{1,2\}$). 
	The size of the network is $N=10,000$, and we set two modules of equal size. 
	$(x_{1},x_{2}) = (1/2,1/2)$ represents the point of a uniform random graph. 
	The average degree of each edge type is $c_{1} = 3$ and $c_{2} = 5$. 
	The shaded region represents an undetectable region, i.e., a region where the inferred module assignments by the EM algorithm are uncorrelated to the planted assignments. 
	The dashed ellipse represents the detectability threshold of Eq.~(\ref{HeimlicherThreshold}). 
	The circles represent the phase boundary obtained through the numerical experiment, where each point represents the average taken over $5$ samples. 
	}
 \label{EMPhaseDiagram-q2}
\end{figure}

When $\Delta c_{\alpha} = 0$ for all $\alpha$, there is no community structure to be detected. 
However, even when a planted partition exists such that $\lvert \Delta c_{\alpha}\rvert > 0$ for some $\alpha$, there is a critical value below which the algorithm cannot detect a partition that is correlated to the planted partition. 
This is called the detectability threshold \cite{Reichardt2008,Decelle2011,Decelle2011a,Mossel2014,Massoulie2014,MooreReview2017,Nadakuditi2012,KawamotoKabashimaEPL2015,CommentDetectabilityThreshold}, and we capture the algorithmic infeasibility from its phase diagram. 

Actually, the detectability threshold of the labeled SBM with BP was previously derived in Ref.~\cite{Heimlicher2012}. 
The threshold is given by 
\begin{align}
\sqrt{\sum_{\alpha>0} \frac{|\Delta c_{\alpha}|^{2}}{P_{\alpha}}} = 2 \sqrt{c}. \label{HeimlicherThreshold}
\end{align}
Notice, however, that this is not the detectability threshold of the EM algorithm.
In Ref.~\cite{Heimlicher2012}, it is assumed that the model parameters are known or learned exactly. 
In practice, we cannot know the planted model parameters a priori, and there is no guarantee that they can be learned accurately. 
Indeed, as shown in Fig.~\ref{EMPhaseDiagram-q2}, the detectability threshold obtained by the EM algorithm (circles) does not coincide with Eq.~(\ref{HeimlicherThreshold}) (dashed ellipse). 
Most importantly, as shown below, Eq.~(\ref{HeimlicherThreshold}) does not exhibit the algorithmic infeasibility. 

To derive the detectability threshold of the EM algorithm, we need to take into account the performance of the M-step. 
We analyze its transient dynamics through the fixed points of Eq.~(\ref{xUpdate}). 
There are three fixed points, and we readily see that two of them are $\hat{x}_{\alpha} = 0$ and $\hat{x}_{\alpha} = 1$. 
For the values of $\{ \psi^{i \to j}_{\sigma} \}$, because we usually have no a priori knowledge, it is common to set uniformly random values as the initial state. 
In this case, because $\bracket{X^{ij}}_{E_{\alpha}} = 1/2$, we have $\hat{x}_{\alpha} = 1/2$ as the third fixed point of Eq.~(\ref{xUpdate}). 
Whereas $\hat{x}_{\alpha} = 0$ and $\hat{x}_{\alpha} = 1$ are unstable fixed points, $\hat{x}_{\alpha} = 1/2$ is a stable fixed point corresponding to the parameter of the uniform random graph. 
Therefore, when $\hat{x}_{\alpha}$ is initially set in the detectable region, it is first attracted toward $1/2$. 
Although the information of the input network flows into the M-step through $\{ \psi^{i \to j}_{\sigma} \}$, this effect is not observable during the transient regime because there is a lag until the structure of the network is reflected in these distributions. 
When the transient regime is over, $\Delta\hat{c}_{\alpha}$ starts to move toward the planted value if it is detectable. 
The robustness of the transient regime is left as an open question here, and our result in the following is conditioned upon the fact that the transient regime exists for sufficiently long steps. 

To analyze the detectability threshold, we analyze the stability of the so-called factorized state in BP, which has $\psi^{i}_{\sigma_{i}} = \psi^{i \to j}_{\sigma_{i}} = 1/2$ for all vertices. 
When BP converges to the factorized state, because we have no information regarding the likely module assignments, we cannot estimate the planted module assignments at better than chance. 
In the case of the standard SBM, the instability condition of BP with respect to the factorized state can be characterized through the eigenvalues of the so-called non-backtracking matrix \cite{Krzakala2013}. 
Herein, we consider the weighted non-backtracking matrix $B^{\prime}$, which is defined as follows. 
This is a $2L \times 2L$ matrix whose element is defined on a pair of directed edges $i\to i^{\prime}$ and $j\to j^{\prime}$, and has  
\begin{align}
& B^{\prime}_{i\to i^{\prime}, j\to j^{\prime}} = \frac{\Delta \hat{c}_{\alpha}}{q c_{\alpha}} \delta_{i j^{\prime}}(1 - \delta_{i^{\prime} j}) 
\hspace{10pt}
\begin{cases}
i\to i^{\prime} \in E, \\
j\to j^{\prime} \in E_{\alpha}
\end{cases}.
\end{align}
Note here that $B^{\prime}$ is a function of $\{\Delta \hat{c}_{\alpha}\}$, which varies dynamically. 
The existence of an eigenvalue $\lambda(B^{\prime})$ of $B^{\prime}$ that satisfies $\lvert\lambda(B^{\prime})\rvert > 1$ yields the instability condition of the factorized state. 
We denote the radius of the spectral band of $B^{\prime}$ as $\lambda_{\mathrm{b}}(B^{\prime})$, and an isolated eigenvalue outside of the spectral band as $\lambda_{\mathrm{iso}}(B^{\prime})$. 

\begin{figure}[t]
 \begin{center}
   \includegraphics[width= \columnwidth]{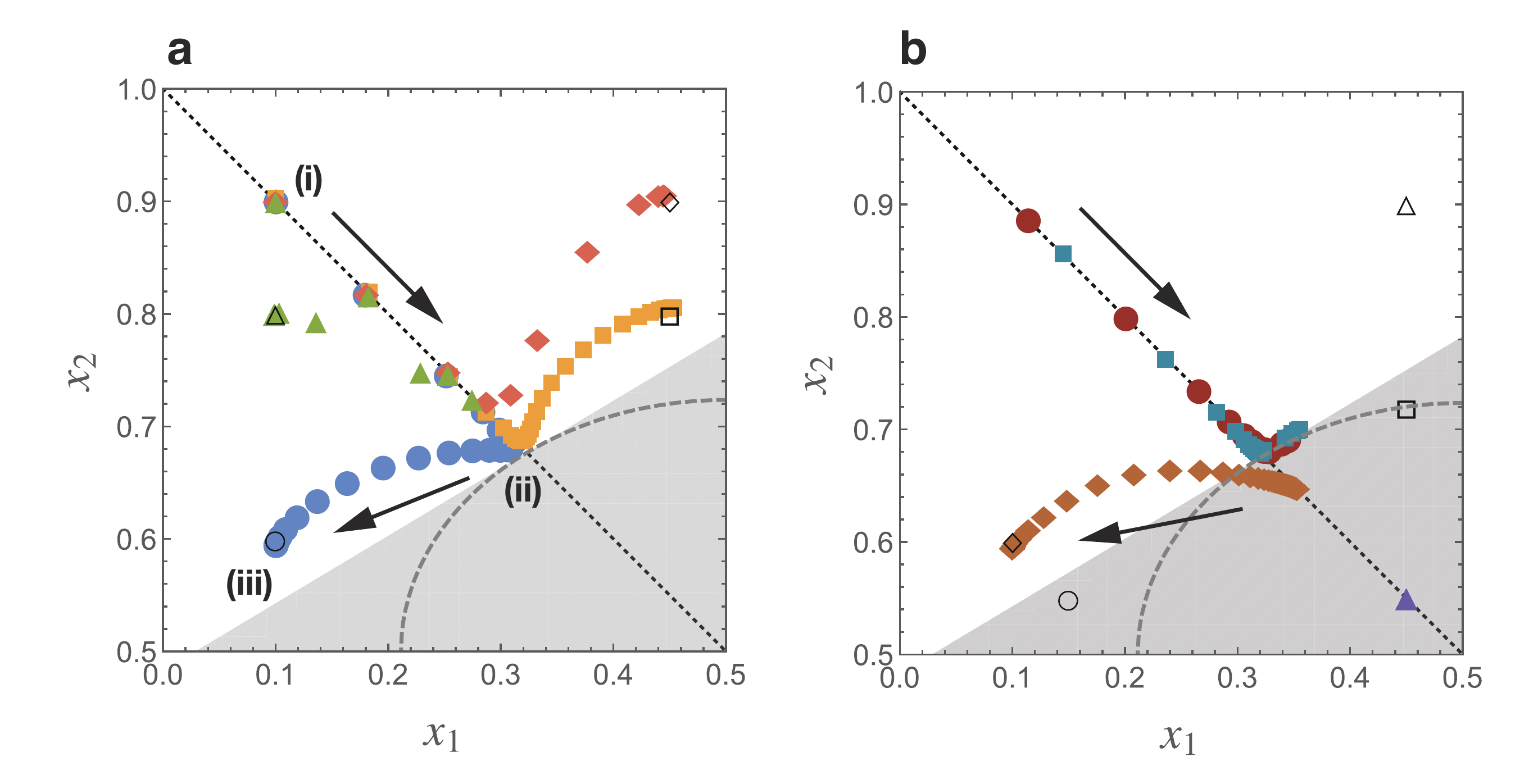}
   \includegraphics[width= \columnwidth]{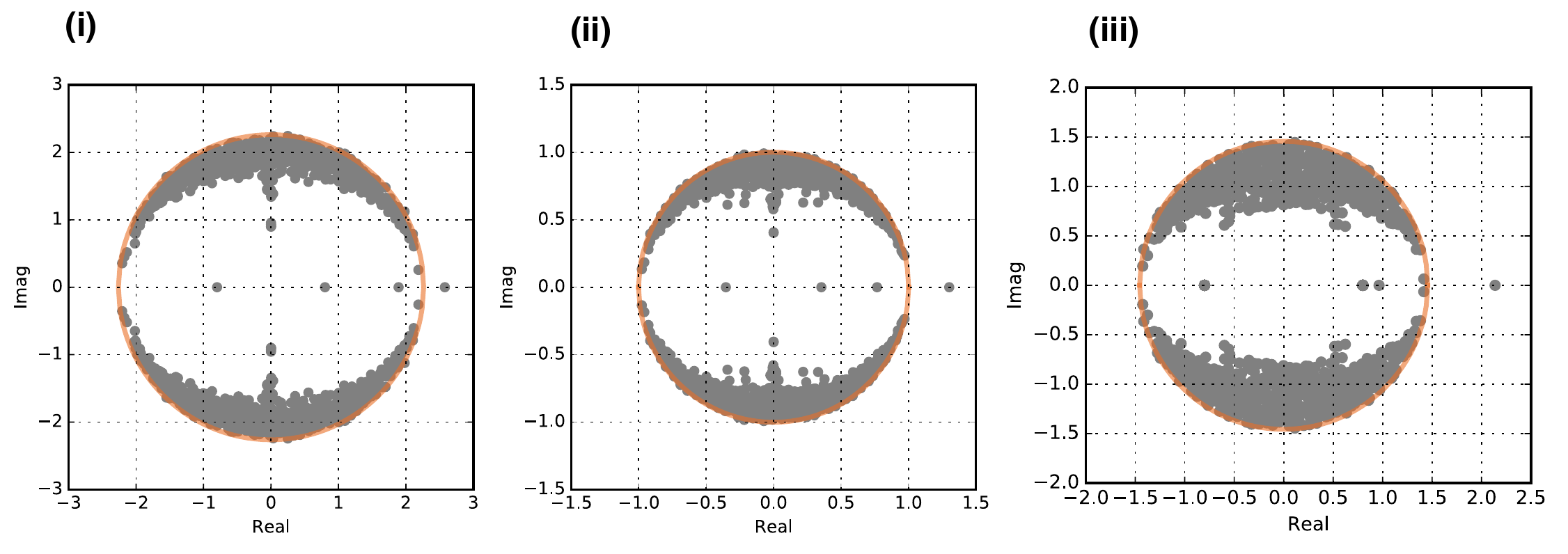}
 \end{center}
 \caption{
	(Color online) 
	(top) Trajectories of parameter learning based on the M-step of the EM algorithm for various planted values of $(x_{1}, x_{2})$. 
	This plot shows the upper-left region of the phase diagram shown in Fig.~\ref{EMPhaseDiagram-q2}, and we consider the same labeled SBM as in Fig.~\ref{EMPhaseDiagram-q2}. 
	The arrows show the directions in which the estimated parameters move. 
	({\bf a}) The trajectories of the estimates $(\hat{x}_{1}, \hat{x}_{2})$ for the case that the planted values $(x_{1}, x_{2})$ (shown in open symbols) are in the detectable region. 
	The trajectory for the graph with the planted value $(x_{1}, x_{2}) = (0.1, 0.6)$ is represented by blue circles; 
	$(x_{1}, x_{2}) = (0.45, 0.8)$ is represented by yellow squares; 
	$(x_{1}, x_{2}) = (0.45, 0.9)$ is represented by red diamonds; and 
	$(x_{1}, x_{2}) = (0.1, 0.8)$ is represented by green triangles, respectively. 
	In all cases, the initial estimate is set as $(\hat{x}_{1}, \hat{x}_{2}) = (0.1, 0.9)$. 
	The dotted line represents the line with slope $-1$. 
	({\bf b}) The trajectories of the estimates $(\hat{x}_{1}, \hat{x}_{2})$ in other cases. 
	The trajectories with the planted values $(x_{1}, x_{2}) = (0.15, 0.55)$ (red circles) and $(x_{1}, x_{2}) = (0.45, 0.72)$ (cyan squares) are the cases in which the planted values (shown in open symbols) are in the undetectable region though the initial estimates are set as $(\hat{x}_{1}, \hat{x}_{2}) = (0.1, 0.9)$. 
	The trajectories with the planted values $(x_{1}, x_{2}) = (0.1, 0.6)$ (orange diamonds) and $(x_{1}, x_{2}) = (0.45, 0.9)$ (purple triangles) are the cases in which $(\hat{x}_{1}, \hat{x}_{2})$ is initially located in the undetectable region ($(\hat{x}_{1}, \hat{x}_{2}) = (0.45, 0.55)$) though the planted values are in the detectable region. 
	(bottom) Spectra of the weighted non-backtracking matrix $B^{\prime}$ in the complex plane with $N=500$ corresponding to ({\bf i}) $(\hat{x}_{1}, \hat{x}_{2}) = (0.1, 0.9)$, ({\bf ii}) $(\hat{x}_{1}, \hat{x}_{2}) = (0.323, 0.677)$, and ({\bf iii}) $(\hat{x}_{1}, \hat{x}_{2}) = (0.1, 0.6)$. 
	The solid line (red) represents the circle with radius $|\lambda_{\mathrm{b}}|$. 
	}
 \label{EMtrajectories-q2}
\end{figure}

The EM algorithm encounters the detectability threshold as follows. 
When the initially set values of $\lvert\Delta\hat{c}_{\alpha}\rvert$ are sufficiently large, the spectral band of $B^{\prime}$ initially has a radius $\lambda_{\mathrm{b}}(B^{\prime})$ exceeds 1 (e.g., Fig.~\ref{EMtrajectories-q2} ({\bf i})), whereas the radius shrinks as $\{\Delta\hat{c}_{\alpha}\}$ is updated during the M-step. 
During this stage, the factorized state is always unstable, and thus the detectability threshold will not be observed. 
Then, when $\lambda_{\mathrm{b}}(B^{\prime})$ reaches unity (e.g., Fig.~\ref{EMtrajectories-q2} ({\bf ii})), as long as an isolated eigenvalue $\lambda_{\mathrm{iso}}(B^{\prime})$ outside of the spectral band is absent, the factorized state becomes stable. 

In summary, the detectability threshold of the EM algorithm is derived as follows. 
We determine the estimate of the model parameters $\{\hat{c}_{\alpha}\}$ at the detectability threshold based on the condition $\lvert\lambda_{\mathrm{b}}(B^{\prime})\rvert = 1$. 
Then, given these estimates, we determine the set of planted values $\{\Delta c_{\alpha}\}$ that satisfies the condition $\lvert\lambda_{\mathrm{iso}}(B^{\prime})\rvert = 1$.

We can derive the radius $\lvert\lambda_{\mathrm{b}}(B^{\prime})\rvert$ of the spectral band by applying the result in Ref.~\cite{NeriMetzPRL2016}. 
We have 
\begin{align}
\lvert\lambda_{\mathrm{b}}(B^{\prime})\rvert &= \frac{1}{2\sqrt{c}} \sqrt{ \sum_{\alpha>0} \frac{|\Delta \hat{c}_{\alpha}|^{2}}{P_{\alpha}} }. 
\label{lambda-b}
\end{align}
When $\Delta c_{\alpha} = \Delta \hat{c}_{\alpha}$ holds for all $\alpha$, the condition $|\lambda_{\mathrm{b}}| = 1$ coincides with Eq.~(\ref{HeimlicherThreshold}).

We then solve for the isolated eigenvalue $\lambda_{\mathrm{iso}}(B^{\prime})$. 
We let $v^{\sigma}_{i \to j}$ be an eigenvector element of $B^{\prime}$, in which the vertex $i$ has the planted module assignment $\sigma$. 
Because the non-backtracking matrix is an oriented matrix, analogous to Ref.~\cite{NeriMetzPRL2016}, $\lambda_{\mathrm{iso}}$ is a solution of the following eigenvalue equation. 
\begin{align}
\lambda_{\mathrm{iso}} u^{\sigma} &= \sum_{\sigma^{\prime}} J_{\sigma \sigma^{\prime}} u^{\sigma^{\prime}}, \label{AveragedEigenvalueEquation}
\end{align}
where $u^{\sigma} = \bracket{v^{\sigma}_{i \to j}}$ is the ensemble average of the eigenvector element, and $J_{\sigma \sigma^{\prime}} \in J$ is defined as 
\begin{align}
J_{\sigma \sigma^{\prime}} &\equiv \frac{1}{q} \sum_{\alpha>0} c^{\alpha}_{\sigma \sigma^{\prime}} \frac{\Delta \hat{c}_{\alpha}}{q c_{\alpha}}. 
\end{align}
Among the eigenvalues of the matrix $J$, the one that corresponds to $\lambda_{\mathrm{iso}}$ is 
\begin{align}
\lambda_{\mathrm{iso}} &= \sum_{\alpha>0} \frac{\Delta c_{\alpha}}{2 \sqrt{c_{\alpha}}} \frac{\Delta \hat{c}_{\alpha}}{2 \sqrt{c_{\alpha}}}. 
\label{lambda-iso}
\end{align}

Using Eqs.~(\ref{lambda-b}) and (\ref{lambda-iso}), we now derive the phase boundary that we numerically obtained through Fig.~\ref{EMPhaseDiagram-q2}. 
We set the initial estimate $(\hat{x}_{1},\hat{x}_{2})$ at near the corner of the $(x_{1}, x_{2})$-plane. 
As can be observed in Fig.~\ref{EMtrajectories-q2} (top), each estimate $\hat{x}_{\alpha}$ is attracted toward the point of the uniform network at an equal rate until it satisfies the condition $\lvert\lambda_{\mathrm{b}}(B^{\prime})\rvert = 1$ (Fig.~\ref{EMtrajectories-q2} ({\bf ii})), or equivalently,  $|\hat{x}_{\alpha} - 1/2| = (2\sqrt{c})^{-1}$ for both $\alpha$. 
Given these estimates, the condition $\lvert\lambda_{\mathrm{iso}}(B^{\prime})\rvert=1$ yields 
\begin{align}
\sum^{p}_{\alpha=1} P_{\alpha} \left| x_{\alpha} - \frac{1}{2} \right| = \frac{1}{2 \sqrt{c}}. \label{AlgoDetectabilitySymInit}
\end{align} 
This is the boundary of the shaded region in Fig.~\ref{EMPhaseDiagram-q2}. 
In terms of $\Delta c_{\alpha}$, Eq.~(\ref{AlgoDetectabilitySymInit}) is $\sum_{\alpha>0} |\Delta c_{\alpha}| = 2\sqrt{c}$. 
As shown in Fig.~\ref{EMtrajectories-q2}b (circles and squares), we can confirm that the planted model parameters in the undetectable region cannot be learned. 

Owing to its relevance regarding the condition of Eq.~(\ref{lambda-b}), the detectability threshold varies depending on the initial estimates of $(\hat{x}_{1},\hat{x}_{2})$. 
When $(\hat{x}_{1},\hat{x}_{2})$ is initially in an undetectable phase, as shown in Fig.~\ref{EMtrajectories-q2}b, the planted model parameters can be learned when the instability condition $\lvert\lambda_{\mathrm{iso}}(B^{\prime})\rvert>1$ is met (diamonds), and cannot be learned otherwise (triangles). 
The phase boundary is generally a simplex tangent to the dashed ellipse.

\begin{figure}[t!]
 \begin{center}
   \includegraphics[width=0.99 \columnwidth]{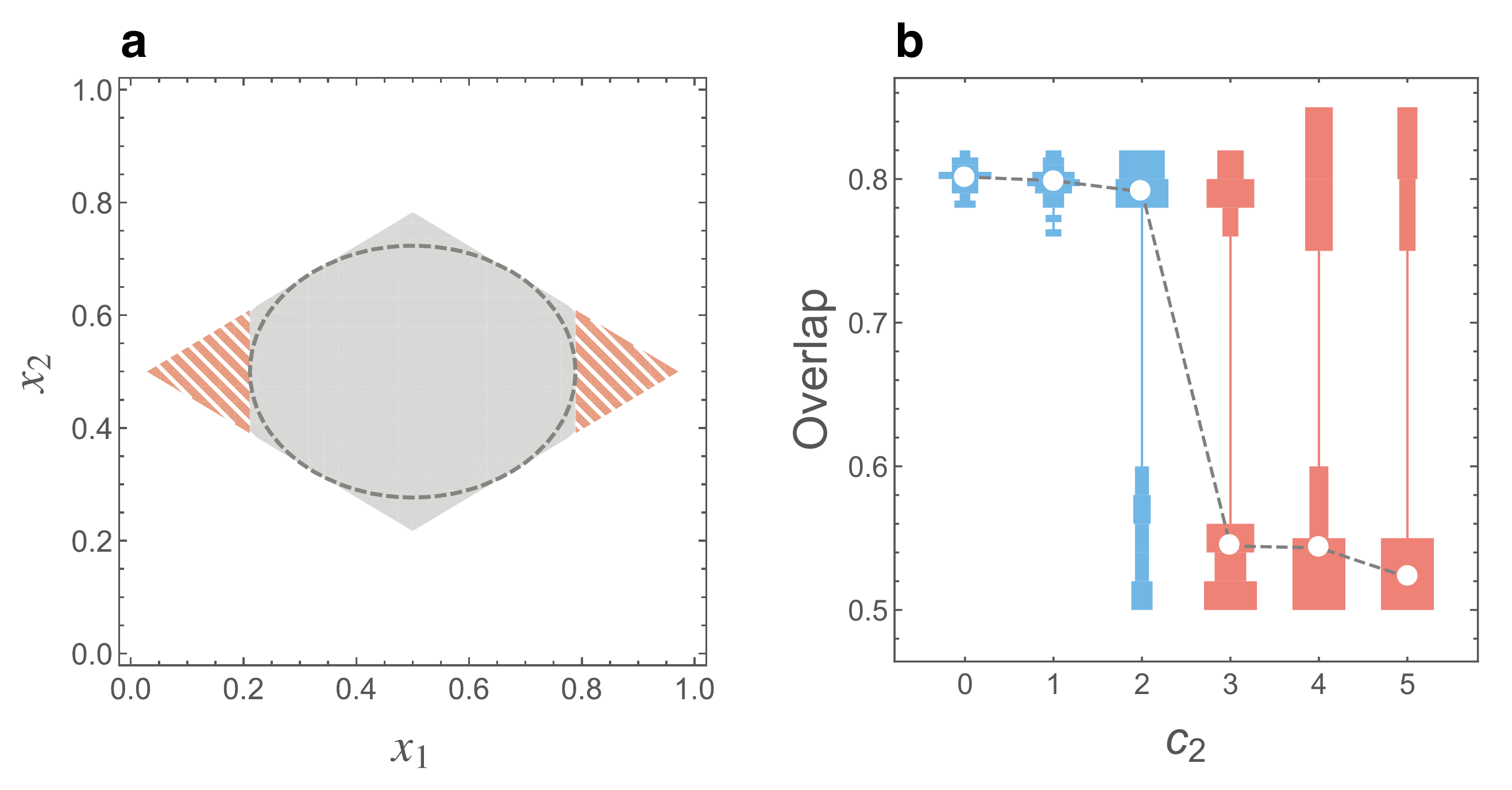}
 \end{center}
 \caption{
	(Color online) ({\bf a}) Same detectability phase-diagram as in Fig.~\ref{EMPhaseDiagram-q2}. 
	The striped region represents the undetectable region that becomes detectable when the edges of $\alpha=2$ are discarded. 
	({\bf b}) Vertical histograms of the overlap distribution. 
	The same labeled SBM as that in Fig.~\ref{InfeasibilityPhaseDiagram-q2}a is considered, and the instances with $N=10,000$, $c_{1}=3$, and $(x_{1}, x_{2}) = (0.85,0.45)$ are generated. 
	The histograms of various values of $c_{2}$ are horizontally aligned (30 samples for each histogram). 
	The ones in the detectable phase, i.e., $c_{2} \le 2$, are indicated in blue, while the ones in the undetectable phase, i.e., $c_{2} \ge 3$, are in red. 
	The white points (connected via dashed lines) indicate the medians of overlaps. 
	The population of the success and failure changes at the critical value that we estimated. 
	}
 \label{InfeasibilityPhaseDiagram-q2}
\end{figure}

Armed with an analytical expression of the detectability phase-diagram, we then ask whether we can make the community structure detectable simply by discarding the edges of one type; the existence of such a phase indicates the algorithmic infeasibility. 
When we completely discard the edges of $\alpha=2$, and utilize only the edges of $\alpha=1$, the detectability region is given by $\lvert \Delta c_{1} \rvert > 2\sqrt{c_{1}}$. 
Thus, in the striped region in Fig.~\ref{InfeasibilityPhaseDiagram-q2}, the lower-order network outperforms the higher-order counterpart. 
This phenomenon is more prominent when the fractions of the edges $\{P_{\alpha}\}$ are very heterogeneous. 
Note that the distinction between Eqs.~(\ref{HeimlicherThreshold}) and (\ref{AlgoDetectabilitySymInit}) is essential. 
Because Eq.~(\ref{HeimlicherThreshold}) (dashed ellipse) is tangent to $\lvert \Delta c_{1} \rvert > 2\sqrt{c_{1}}$, the algorithmic infeasibility will not be observed when the model parameters are known exactly.

Our detectability analysis of the EM algorithm established the algorithmic infeasibility of community detection theoretically. 
The results here offer an incentive to discard some of the information in higher-order networks for a better community detection performance. 
Although we focused on a simple community structure of two equally sized modules, our analysis can be greatly generalized \cite{ADT-Full}.

\textit{Acknowledgments}--- The author thanks Yoshiyuki Kabashima, Tomoyuki Obuchi, and Jean-Gabriel Young for fruitful discussions and valuable comments. This work was supported by the New Energy and Industrial Technology Development Organization (NEDO).


\bibliographystyle{apsrev}
\bibliography{bib-EMdetectability}

\begin{thebibliography}{25}
\bibitem[{Com()}]{CommentDetectabilityThreshold}
\bibinfo{howpublished}{The detectability threshold is usually defined as the
  fundamental limit where a network is indistinguishable from an instance of
  the uniform random graph. Here we define it as the threshold of the
  algorithm.}
\end{thebibliography}

\end{document}